\documentclass[11pt]{article}
\usepackage{aaspp4,eqsecnum,astrobib,graphicx}

\begin{document}

\def\Ho{\ifmmode {\rm\,H_0}\else ${\rm\,H_0}$\fi}
\def\hnot{\ifmmode {\rm\,H_0}\else ${\rm\,H_0}$\fi}
\def\h0{\ifmmode {\rm\,H_0}\else ${\rm\,H_0}$\fi}
\def\hnotunit{\ifmmode {\rm\,km\,s^{-1}\,Mpc^{-1}}\else
    ${\rm\,km\,s^{-1}\,Mpc^{-1}}$\fi}
\def\qnot{\ifmmode {\rm\,q_0}\else ${\rm q_0}$\fi}
\def\q0{\ifmmode {\rm\,q_0}\else ${\rm q_0}$\fi}
\def\msolar{M_\odot}
\def\lya{Ly$\alpha$}
\def\ciii{\ifmmode {\mbox{C~{\sc iii}}} \else C~{\sc iii}\fi}
\def\ciiil{C~{\sc iii}]\ }
\def\civ{\ifmmode {\mbox{C~{\sc iv}}} \else C~{\sc iv}\fi}
\def\civll{C~{\sc iv}$\lambda\lambda1548,50$}
\def\ergps{{\rm erg \> s^{-1}}}

\begin{flushright}
\today
\end{flushright}

\title{Jet-Induced Emission-Line Nebulosity and  Star Formation in the
High-Redshift Radio Galaxy  4C41.17}
 
\bigskip
\author{Geoffrey V. Bicknell$^1$}
\author{Ralph S. Sutherland$^1$}
\author{Wil J. M. van Breugel$^2$}
\author{Michael A. Dopita$^1$}
\author{Arjun Dey$^3$ }
\author{George K. Miley$^4$}

\vspace {33 pt}

\begin{enumerate}
\item ANU Astrophysical Theory Centre, Research School of Astronomy \& Astrophysics,
Australian National University. Postal address: Mt Stromlo Observatory, Private Bag, Weston
PO, ACT, 2611, Australia. Email addresses: Geoffrey Bicknell: Geoff.Bicknell@anu.edu.au;
Ralph Sutherland: ralph@mso.anu.edu.au; Michael Dopita: mad@mso.anu.edu.au
\item Institute of Geophysics \& Planetary Physics, LLNL, Livermore, CA 94550. Email:
wil@igpp.llnl.gov
\item KPNO/NOAO, 950 N. Cherry Ave., PO Box 26732, Tucson, AZ 85726. Present
address: Dept. of Physics and Astronomy, The Johns Hopkins University, Baltimore, MD 21218.
Email: dey@skysrv.pha.jhu.edu
\item Leiden Observatory, PO Box 9513, 2300 RA, Leiden, The Netherlands. Email:
miley@strw.leidenuniv.nl
\end{enumerate}

\newpage

\begin{abstract} The high redshift radio galaxy 4C41.17 has been shown in
earlier work to consist of a powerful radio source in which there is strong
evidence for jet-induced star formation along the radio axis. We argue that
nuclear  photoionization is not responsible for the excitation of the emission
line clouds along the axis of the radio source and we therefore construct a
jet-cloud interaction model to explain the major features revealed by the
detailed radio, optical and spectroscopic data of 4C41.17. The interaction of a
high-powered ($\sim 10^{46}
\> \rm ergs \> s^{-1}$)  jet with a dense cloud in the halo of 4C41.17
produces shock-excited emission-line nebulosity through $\sim 1000 \> \rm km
\> s^{-1}$ shocks and induces star formation. The \ciiil to \civ\ line ratio
and the \civ\ luminosity emanating from the shock, imply that the pre-shock
density in the line-emitting cloud is high enough ($\hbox{hydrogen density}
\sim 1-10 \> \rm cm^{-3}$) that shock initiated star formation  could proceed
on a timescale ($\sim
\hbox {a few} \times  10^6
\> \rm yrs$), well within the estimated dynamical age ($\sim 3 \times 10^7 \> \rm yrs$)
of the radio source.  The star formation efficiency in the shocked cloud is 
$\sim 1 \%$. Broad (FWHM
$\approx 1100 - 1400\> \rm km \> s^{-1}$) emission lines are attributed to the
disturbance of the gas cloud by a partial bow--shock and narrow emission lines
(FWHM 
$\approx 500 - 650 \> \rm km \> s^{-1}$ ) (in particular \civll) arise in
precursor emission in relatively low metallicity gas. 

The implied baryonic mass $\sim 8 \times 10^{10}\>\msolar$  of the cloud is
high and implies that Milky Way size condensations existed in the environments 
of forming radio galaxies at
a redshift of 3.8. Our interpretation of the data provides a physical basis
for the alignment of the radio, emission-line and UV continuum images in some
of the highest redshift radio galaxies and the analysis presented here may form a
basis for the calculation of densities and cloud masses in other high redshift
radio galaxies. 
 
\end{abstract}
 
\keywords{galaxies: active --- galaxies: galaxies --- elliptical:
high-redshift --- radio continuum: galaxies}
 
\newpage

\section {Introduction}
\label{s:intro}

One of the most intriguing discoveries in the study of high-redshift radio
galaxies (HzRG) has been that the rest-frame UV continuum emission from  their
parent galaxies is aligned with the non-thermal radio emission 
\cite{mccarthy87a,chambers87a}. The nature of this continuum and `alignment
effect' has remained  unclear. In nearby radio galaxies  evidence has been
found for jet-induced star formation, scattered light from  hidden quasar-like
AGN  and nebular re-combination continuum 
(\citeN{vanbreugel85a};\citeN{vanbreugel93a};\citeN{dey96a};\citeN{tadhunter96a};\citeN{dickson95a};
\shortciteN{alighieri89a};\citeN{cimatti96a}). A good example of radio-aligned
UV emission in a very high-redshift radio galaxy is 4C41.17 at
$z = 3.800$, which has been extensively studied at optical and radio
wavelengths 
\cite{chambers90a,miley92a,carilli94a,chambers96b}. Recent HST observations
have shown that the rest-frame UV 
morphology of 4C41.17 consists of four main regions, the brightest of which
(4C41.17-NE) contains an edge-brightened bifurcated feature consisting of 
several compact knots located between the radio nucleus and a bright radio
knot 
\cite{vanbreugel98a}. 

Deep spectropolarimetric observations with the W. M. Keck Telescope by
\citeN{dey97a} show that 4C41.17 is unpolarized between
$\lambda_{\rm rest} \sim 1400 \> \AA - 2000 \> \AA $,  implying that scattered
light does not dominate the aligned UV continuum. Instead, the observations show
absorption lines and P-Cygni-like features that are similar to those seen in $z
\approx 2 - 3$ star forming galaxies and nearby Wolf-Rayet starburst systems.  The
possibility of jet-induced star formation in 4C41.17 and other HzRGs has been
suggested before \cite{deyoung81a,deyoung89a,rees89a,begelman89a,chambers90a,daly90a}
but until now has lacked sufficient observational basis. In this paper we revisit the
jet-induced star formation scenario for 4C41.17 in the light of the new data that are
now available, and present a self-consistent model in which interactions between jets
and dense clouds in  4C41.17 produce both shock-excited line emission and induce star
formation. As we show below, it is fortunate that both phenomena occur since
information provided by the former process enables us to better constrain the
parameters relating to the latter.

Throughout this paper we assume that \hnot=50\hnotunit\ and \qnot=0.1. The
luminosity distance d$_L$, angular size distance d$_A$ and linear scale at the
redshift of 4C41.17 ($z=3.800$) are then 51.6 Gpc, 2.24 Gpc, and 10.8 $\rm kpc
\> \hbox{arcsec}^{-1}$ respectively. We follow the notation of 
\citeN{chambers90a} and \citeN{carilli94a} in referring to the radio features
(components, knots etc.).

\section {HST Observations of 4C41.17 and the Relationship to the Radio
Emission}
\label{s:summary}

The details of HST imaging of 4C41.17 are given in \shortciteN{vanbreugel98a}.
Here, we summarize some of the pertinent details of these images and their
relationship to the radio emission in order to facilitate the following
theoretical discussion.

The montage in Figure~1 shows three HST images in different
bands with the X-band radio images of \citeN{carilli94a} superimposed in the
form of contours. The top image is a deep rest-frame UV image (F702W filter,
$\lambda{_{rest}}$
$\sim$ 1430\AA; 6.0 hours exposure); the middle image was acquired through the
F569W filter, which includes
\lya (2.0 hours exposure); the bottom image is a \lya\ image (LRF filter at $\lambda_c \sim$
5830 \AA; 2.0 hours exposure). All of these images show strongly aligned
non-thermal and thermal components. The direct association of the radio
components with {\em both} UV continuum and \lya\ emission, together with 
the spectroscopic 
evidence for young stars from the Keck observations, strongly points to
jet-induced star formation. In particular, the radio knot B2 (the second from
the left in these images) is associated with the brightest \lya\ region and
the F702W and F569W images reveal an interesting bifurcated or oval feature
(approximated with a 0.8 by $0.24 \arcsec$ [ $\sim 8.6 \times  2.6$~kpc] oval or
parabola, shown enlarged on the right of Figure~1, which we interpret
as tracing the locus of newly formed stars. See \shortciteN{vanbreugel98a} for a
more detailed discussion of these images.

In Figure~2 a $0.3 \arcsec$ smoothed version of the F702W
image is displayed. This brings out an additional star forming region to the
South  of the regions evident in the unsmoothed version.
\shortciteN{vanbreugel98a} have estimated the star-formation rates in these
regions from the UV luminosity, using the relationship between ultraviolet
flux and star formation rate determined by \citeN{conti96a}. The estimated
star formation rates in the various regions are given in table~1.

Following \shortciteN{vanbreugel98a} we adopt the following nomenclature for
the components in the HST image: The NE component is the region of
edge-brightened UV emission located on the core side of the bright radio knot
B2; NEE is the more diffuse component to the East of this. NW is the UV
component along the radio axis on the Western side of the radio core and S
represents is the clumpy component to the South, revealed by the smoothed
image.

The evidence for jet-induced line emission and star formation in the brightest
UV emission region in 4C41.17 (4C41.17NE) is compelling and can be summarized
as follows \cite{vanbreugel98a,dey97a}:

\begin{itemize}

\item The star formation rate per square kiloparsec in the four UV
bright regions mentioned in the introduction is by far the greatest
in 4C41.17NE \shortcite{vanbreugel98a}.  (1996).  The morphology of
4C41.17NE and its close proximity to the radio knot, B2, strongly
indicate that star formation has been induced by the interaction
between the northern jet of the radio source and the cloudy medium of
the forming parent galaxy as expected in jet-induced star formation
models  (e.g. \citeN{deyoung89a}).  The random distribution and lower
star formation rates in the 4C41.17S knots, which are comparable to
hose of `Lyman-break' galaxies \shortcite{steidel96a}, suggests that
star formation here is unaided by bowshocks from the radio jet.

\item The HST Ly$\alpha$ image shows a bright arc-shaped feature near B2 at
the apex of the edge-brightened UV structure, suggestive of a strong shock at
a location where the jet interacts with dense ambient gas. Such emission-line
features near bright radio structures are also often seen  in nearby radio
galaxies \cite{vanbreugel85,tadhunter94a} and these have a similar interpretation.

\item  The kinematics of the Ly$\alpha$ emission is very much disturbed in the
aligned component with velocities with respect to systematic $\sim 500-1400\>
\rm km \> s^{-1}$
 and velocity dispersions $\sigma \sim 300 - 600  \> \rm km \> s^{-1}$ 
\cite{dey97a,chambers90a}, suggesting large (projected) velocities and strong
turbulence caused by jet/cloud interactions. 

It follows from the above three points that the emission lines from this
galaxy are probably related either to the star-forming region or to emission
from radiative cloud shocks rather than excitation by UV-X-ray emission from
the active nucleus.

\item The Keck spectra by \citeN{dey97a} show that emission-line gas
associated with the  components B1, B2 and B3 of the inner radio source
4C41.17 consists of two distinct kinematic components: relatively narrow lines
for all species (Ly$\alpha$, N~{\sc v}, Si~{\sc ii}, Si~{\sc iv}, \civ, He~{\sc
ii}, and \ciii) with FWHM 
$\approx 500 - 650 \> \rm km \> s^{-1}$ ($\sigma \approx 220 - 270 \> \rm km
\> s^{-1}$), and broad Ly$\alpha$ and \ciiil with  FWHM
$\approx 1100 - 1400\> \rm km \> s^{-1}$ ($\sigma \approx 470 - 600 \> \rm km
\> s^{-1}$).  (Si~{\sc iv} is possibly broad also; however the estimate of the
line width is complicated by associated absorption.) We assume that the narrow
velocity components are related to the jet-driven radiative shocks, and the
broad components by filaments pulled out of the cloud through the action of
the Kelvin-Helmholtz instability at the jet-cloud interface. This is discussed
further in
\S~\ref{s:interaction}.

\item  The brightest Ly$\alpha$ emission is found on the same side (East)
which has the outer hotspot of 4C41.17 closest the nucleus. This agrees with
the general radio / EELR morphological asymmetry correlation seen in powerful
FR-II radio galaxies \cite{mccarthy91a}, and suggests that the radio source
has been impeded in this direction as a result of its encounter with
relatively dense gas.

\item As we noted in the introduction, the absence of any evidence for a
polarized, scattered AGN continuum supports the notion that,
in the case of 4C41.17, the active nucleus is not responsible for the
extended UV emission.

\end{itemize}

\section{Interaction of Clouds in the Ly$\alpha$ Halo with the  Radio Jets}
\label{s:interaction}

4C41.17 is located at the center of a large Ly$\alpha$ halo
\cite{chambers90a}. The passage of  relativistic jets through such a halo will
inevitably result in substantial jet--cloud interactions. In the case of the
jet-cloud interaction evident near the radio knot B2 we suggest that a
glancing incidence of the jet on the cloud causes a partial bow-shock to be
driven in to the cloud. This is manifest through the associated shock-excited
line emission and associated star formation in the bifurcated structure
referred to above. The jet deposits much of its momentum at this site and it
continues onward to the knot B3 where the decelerated jet plasma accumulates
as a radio ``lobe''. In this section we estimate physical cloud and jet parameters
implied by this interaction model and then consider other emission regions in
the HST images.

An important feature of the spectroscopic observations of 4C41.17 is the three
kinematically distinct components  namely the broad emission lines ($\sigma
\sim 470-600 \> \rm km \> s^{-1}$),  the narrow emission lines ($\sigma \sim
220-270 \> \rm km \> s^{-1}$)  and the narrow absorption lines ($\sigma
\sim 170-340 \> \rm km \> s^{-1}$).  We suggest that these components arise in
the following way (see Figure~4). The narrow emission lines
have a velocity dispersion similar to the halo and are formed either by
locally induced photoionization of halo gas or in the winds of newly formed
stars. The natural location for the narrow absorption lines is in the
atmospheres of the young stars and in some cases the narrow absorption and
emission lines comprise a typical P-cygni-like profile characteristic of
winds from young stars see \citeN{dey97a}. We suggest that the broad emission
lines arise from shock-excited gas which has been significantly disturbed by
the jet-cloud bow shock. This phenomenon is also observed in low redshift
radio galaxies 
\cite{vanbreugel85,tadhunter91a}.

Many of the observed narrow emission lines could be produced either in the
shock or in the photoionized winds of the newly formed stars. An important
exception is \civ\ which is weak in stars older than $3\times 10^6
\> \rm yrs$ \cite{leitherer95a}. Moreover, when this line {\em is} present in
emission in young stars, its strength is comparable to the absorption
strength. In 4C41.17 the \civ\ emission line strength dwarfs the absorption
component and we therefore completely attribute this component of emission to
the effects of the radiative cloud shock. The \civ emission is narrow and this
is a strong indication that most of the flux from this line originates in the
precursor material ahead of the cloud bow shock. As we show below this is
consistent with the velocity $\sim 1000 \> \rm km \> s^{-1}$ that we adopt for
the normal component of this shock. 

\subsection{The Jet Bow Shock}

Let us now focus on the brightest emission line knot adjacent to knot B2 in
the radio image. If a large fraction of the jet momentum flux, $F_{\rm p}$ is
absorbed at this interaction site, and the cross-sectional area of the
(presumably jittering) jet over which the momentum is spread is $A_{\rm jet}$,
then the velocity of the bow shock driven into the cloud of density $\rho_{\rm
cl}$ is given by
$v_{\rm sh} \approx
\left(F_{\rm p}/\rho_{\rm cl} A_{\rm jet}\right)^{1/2}$. For a relativistic
jet, the energy flux,
$F_{\rm E} = cF_{\rm p}$; for a non relativistic jet, $F_{\rm E} \approx
v_{\rm jet}/2 F_{\rm p}$. The FWHM of knot B2 $\approx 0.11\arcsec$
\cite{chambers90a} and this angular scale provides an upper limit for $A_{\rm
jet} \approx 1.1\times 10^{43} \> \rm cm^2$ since the radio emission resulting
from the jet burrowing into the cloud emanates from a larger volume than just
the head of the locally produced radio cocoon. We therefore obtain the
corresponding lower limits for the bow shock velocity: 
\begin{equation}
\begin{array}{r c l l l} v_{\rm sh} &\ga& 1100 \, F_{\rm E,46}^{1/2} \, n_{\rm
H}^{-1/2}  & {\rm km \> s^{-1}} & \hbox{(relativistic)} \\ v_{\rm sh} &\ga&
1600 \, F_{\rm E,46}^{1/2} \beta_{\rm jet}^{-1/2} \, n_{\rm H}^{-1/2}  & {\rm
km \> s^{-1}} & \hbox{(non-relativistic)}
\end{array}
\label{e:bow-shock}
\end{equation} where $10^{46} \, F_{\rm E,46} \> \rm ergs \> s^{-1}$ is the
jet energy flux and
$n_{\rm H} \> \rm cm^{-3}$ is the Hydrogen density in the cloud. These
velocity limits are to be compared to the velocity with respect to systemic
$\sim 500-1400\> \rm km \> s^{-1}$ of \lya\ in this region and also the FWHM of
\lya\ $\sim 700-1400 \> \rm km \> s^{-1}$. With radiative shocks, comparable
fluxes of \lya\ are emitted from both precursor and shocked regions so that
these two measures of the \lya\ velocity field give us a good indication that
the normal component of the bow-shock velocity is of order
$1000 \> \rm km \> s^{-1}$.

\subsection{Shock-excited line emission}

Let us now consider the emission line fluxes and how these relate to the
ambient density and shock velocities. In so doing we are utlising data with quite different
spatial resolutions, the Keck spectra and the HST line and continuum images. 
The emission line components revealed by the Keck spectra are not resolved by the Keck
spectra and we would expect contributions to the emission line luminosity from a number of the
\lya\ emitting regions. However, we expect the most significant contribution to come
from the brightest \lya\ emitting region in the immediate vicinity of the radio knot B2.
Moreoever, the stellar features in the Keck spectra orginate from young stars and the site of
these is revealed by the HST UV continuum images. Therefore, there is good reason to believe
that the Keck spectra relate to the jet-cloud interaction near radio component B2.

 The strongest emission lines in the
spectrum are \lya\ and \civll. As is well known, the transfer of \lya\ is
subject to strong resonant scattering effects making it difficult to directly
infer shock parameters from emission line fluxes. \civ\ is not as strongly
affected by resonant scattering, and we therefore use the luminosity of the
\civ\ doublet to constrain shock parameters. As we have shown above, the
favored velocity of the normal component of the bow shock velocity  is in the
vicinity of $1000 \> \rm km \> s^{-1}$ and it is useful to note that is that
this is compatible with a number of features of the emission. First, the \civ\
emission from a shock with this velocity is dominated by the precursor and the
velocity dispersion of \civ\ ($\sim 250 \> \rm km \> s^{-1}$ is in the range
of halo velocity dispersions is consistent with precursor dominated emission.
Second, the HST \lya\ image shows \lya\ emission ahead of the presumed
location of bow-shock (knot B2) indicating significant precursor emission in \lya. 

The emission from shocks with velocities $\leq 500 \> \rm km \> s^{-1}$ were
calculated for solar metallicities by \citeN{dopita96a} and \citeN{dopita96b}.
Recent work \cite{sutherland99a} extends these calculations to higher
velocities ($\sim 800-900 \> \rm km \> s^{-1}$) and lower metallicities.
Extension to even higher velocities $\ga 1000 \> \rm km \> s^{-1}$ is
difficult at present. However, the results from the \shortciteN{sutherland99a}
calculations certainly give one an indication of the magnitude of shock
emission and the trend with increasing velocity.

For a shock of area
$A_{\rm sh}$, proceeding into gas with pre-shock Hydrogen density, $n_{\rm H}$
we represent the shock luminosity in \civ, $L(\civ)$,  in the form
\begin{equation} 
L(\civ) = \alpha(\hbox{\civ}) \, n_{\rm H} \, A_{\rm sh}
\end{equation} 
The shock coefficient $\alpha(\civ)$ is shown as a function of
shock velocity and metallicity in Figure~3.  The earlier 
Dopita and Sutherland solar shock grid produces \civ\ emission
with a coefficient $\alpha(\civ)$ ranging between $6\times 10^{-5}$ and $5\times 10^{-4}$,
with the dominant source of \civ\ emission coming from the shock structure 
itself.   Extrapolation to $1000 \> \rm km \> s^{-1}$ would give $\alpha(\civ)$
just over $1\times 10^{-3}$.  The extended grid of  \cite{sutherland99a} for higher
velocities and lower metallicities shows that \civ\ emission is much more efficient
than extrapolated from the earlier low velocity grid, as \civ\ emission from the 
precursor region comes to dominate.  At low metallicities the \civ\ emission
in further enhanced, as the temperature of the precursor increases and
\civll\ is excited more efficiently.  For abundances found in the
Magellanic Clouds (LMC \& SMC,\citeN{russell89}), the $\alpha(\civ)$
coefficient grows rapidly to values around 0.01.   Solar and SMC abundances also rise to
similar values and it is not until
$\rm [Fe/H]$  falls to $-2.0$ or lower that the efficiency finally falls below this.

Therefore, for  shock velocities
$\ga 600 \> \rm km \> s^{-1}$ and for a range of metallicities, particularly
LMC metallicities, $\alpha(\hbox{\civ}) \sim 0.01$ so that we adopt this as
a fiducial value.  Also, as we have noted above, the shock emission in
this case is dominated by the emission from the precursor as distinct from
the low velocity
case where the emission is dominated by the shocked gas. This feature of
shock-induced
emission is consistent with the observed velocity dispersion. The other
important point to note is that because of the timescales involved ($\sim 10^7
\> \rm yr$), shocks with the velocities which are relevant here, are fully
radiative.  For the the velocity range of $700-900 \> \rm km \> s^{-1}$ at
LMC abundances, the cooling timescales are $(9.4 - 16.3)\times 10^5
(1.0/n_{\rm H})$yrs,
so that the shocks are fully radiative on timescales short compared to
the source timescale for densities $\ga 1.0 \> \rm cm^{-3}$.

As a fiducial value for the shock area we take the projected area, 
$A_{\rm p} \approx 35 \> \rm kpc^2$, obtained by counting HST pixels in the 
F569W image in the NE region, above a 'sky' value, it is about 50\% more than
the area of the oval area the F702W and F569W images in
Figure~1 and is equivalent to the dashed rectangle in the centre
right panel.  With this fiducial value for
$A_{\rm sh}$ the predicted \civ\ luminosity from the shock is 
\begin{equation} L({\rm \civ}) \approx 3 \times 10^{42} \left( \frac
{\alpha({\rm \civ})}{0.01} \right) \, n_{\rm H} \, \left( \frac{A_{\rm
sh}}{A_{\rm p}} \right)^{-1} 
\> \rm ergs \> s^{-1}
\end{equation} where the value of $\alpha({\rm \civ})$ is really an average of
the different normal components of velocity over the bow-shock surface. 
Comparing the predicted \civ\ luminosity with that observed,
$\approx 4.2 \times 10^{43} \> \rm ergs \> s^{-1}$, one can see that, if a
fraction $f({\rm \civ})$ the \civ\ luminosity emanates from this region, then
$n_{\rm H} \sim 10 f({\rm \civ}) \, \rm cm^{-3}$. 

By way of the lower limits for the bow shock velocity [see
equations(\ref{e:bow-shock})] upper limits on the energy flux can be estimated
from:
\begin{equation}
\begin{array}{r c l l} F_{\rm E,46} & \la &  0.77 \, n_{\rm H} \, \left( \frac
{v_{\rm sh}}{10^3 \> \rm km \> s^{-1}} \right)^2 \> 
\rm ergs \> s^{-1} & (\hbox{relativistic}) \\ F_{\rm E,46} & \la & 0.39 \,
n_{\rm H} \, \beta_{\rm jet} \, 
\left( \frac {v_{\rm sh}}{10^3 \> \rm km \> s^{-1}} \right)^2 \> \rm ergs \>
s^{-1} & (\hbox{non-relativistic}) 
\end{array}
\label{e:fe}
\end{equation} Taking into account the likely range of number densities and
the range of velocities in the shocked material, it is evident that the upper
limit on the jet energy flux is of order $10^{47} \> \rm ergs
\> s^{-1}$ for a relativistic jet. For $\beta_{\rm jet} \approx 0.1$, the
upper limit
 is of order $10^{46} \> \rm ergs \> s^{-1}$. However, note that a jet with
$\beta_{\rm jet} \sim 0.1$ is unlikely to be supersonic. We expect that all
jets in such sources are initially relativistic and the critical velocity at
which they become subsonic is approximately $0.3 \, c$
\cite{bicknell94a}.

\subsection{Star formation in the shocked cloud}

Star formation initiated by shocks has been a process that has been considered
in many contexts for some time and much of the theoretical underpinnings of
the subject were treated in a fundamental paper by \citeN{elmegreen78a}. They
consider a shocked layer of surface density $\sigma$ confined by both self
gravity and an external pressure $P_{\rm ext}$. The reader is referred to
Figure~5 for a description of the shock geometry. The Elmegreen and Elmegreen
analysis involves the parameter
\begin{equation} A = \left[ 1 + \frac {2P_{\rm ext}}{\pi G \sigma^2}
\right]^{-1/2}
\label{e:A}
\end{equation} where $G$ is the constant of gravitation. Such a layer is
strongly self-gravitating when $A
\rightarrow 1$; the confinement is dominated by external pressure when $A
\rightarrow 0$. With
$\rho_{00}$ as the density at the layer midplane, and $H$ as  the
half-thickness of the layer, the temporal frequency
$\omega$ and the  wavelength, $\lambda$, of a perturbation are given in terms
of their non-dimensional values,
$\Omega$ and $\nu$, by
\begin{equation}
\omega = \left( 4 \pi G \rho_{00} \right)^{1/2} \Omega \qquad\hbox{and} \qquad
\lambda = 2\pi H \nu^{-1}
\end{equation}

In order that jet-induced star formation be effective, the time-scale for
gravitational instability should be less than (and preferably much less than)
the dynamical time-scale $\sim 10^7 \> \rm yr$ for the shocked region that we
are considering. We therefore adopt a fiducial timescale of $10^6 \> \rm yr$
for gravitational instability. The external pressure confining the
recombination layer following a strong radiative shock into material with a
pre-shock density $\rho$, is $P_{\rm ext} \approx \rho v_{\rm sh}^2$ and the
surface density of the accreting layer is $\sigma = \rho v_{\rm sh} t$. Hence,
the instability parameter defined by equation~(\ref{e:A}) above is given by
\begin{equation} A = \left[ 1 + 4.0 \times 10^3 n_H^{-1} t_6^{-2}
\right]^{-1/2} 
\approx 1.6 \times 10^{-2} \, n_{\rm H}^{1/2} t_6
\end{equation} This places the layer in the regime where it is not dominated
by self-gravity. Interestingly, however, fragmentation on a short enough
timescale can occur. This is revealed by the following estimates of the
instability timescale, the length scale corresponding to the maximal growth
rate and the minimal length scale on which instability will occur.
Fragmentation in this parameter regime  has also been pointed out by
\citeN{whitworth94a}.

For $A << 1$, \citeN{elmegreen78a} numerically estimate the nondimensional
values of the maximal growth rate,
$\Omega_{\rm mgr}$, the wave number for maximal growth, $\nu_{\rm mgr}$, and
the maximum wave number for instability, $\nu_c$:
\begin{equation} -\Omega_{\rm mgr}^2 \doteq 0.139 \qquad A \nu_{\rm mgr}
\doteq 0.294 \qquad A \nu_c \doteq 0.639
\end{equation} These non-dimensional values can be converted to physical units
using the above relations between the half-thickness of the layer, the surface
density and the central density, $HA = 2^{-1} \sigma
\rho_{00}^{-1}$, resulting in the following expressions for the timescale of
maximum growth,
$t_{\rm mgr}$, the wavelength, $\lambda_{\rm mgr}$ of the maximally growing
perturbation and the minimum wavelength for instability, $\lambda_{\rm min}$:
\begin{eqnarray} t_{\rm mgr} &\doteq& 58 \, \left( \frac {\rho}{\rho_{\rm
00}}  \right)^{1/2} \, n_{\rm H}^{-1/2} \>
\rm Myr \\
\lambda_{\rm mgr} &\doteq& 11 \, \left( \frac {\rho}{\rho_{00}} \right) \, 
\left( \frac {v_{\rm sh}}{10^3 \> \rm km  \> s^{-1}}  \right) \, t_6 \> \rm
kpc \\
\lambda_{\rm min} &\doteq& 5.1 \> \left( \frac {\rho}{\rho_{00}} \right) \, 
\left( \frac {v_{\rm sh}}{10^3 \> \rm km  \> s^{-1}}  \right) \, t_6 \> \rm
kpc 
\end{eqnarray} The relevance of these estimates to the present situation is
realized when we allow for the fact that in a radiative shock the ratio,
$\rho_{00}/\rho$, of recombination to pre-shock densities is typically of order
100. 
For $\rho_{00}/\rho \sim 100$, $t_{\rm mgr} \approx 6-2 \> \rm Myr$ for $n_{\rm H} = 1-10 \>
\rm cm^{-3}$. The linear scale corresponding to the maximum growth rate and the minimum growth
scale both increase with time and indicate that the sizes of the star
formation regions are of order a few hundred parsecs after about $3 \times 10^6 \> \rm yr$.

Thus, for the range of pre-shock densities, $n_{\rm H} \sim 1-10 \> \rm
cm^{-3}$ that we have identified from the shock dynamics, it is quite clear
that gravitational instability occurs on timescales comfortably within the
dynamical timescale of the jet-cloud interaction. Moreover, the sizes of the
star formation regions fit well within the structures observed at the 
interaction site. Indeed, it is interesting to note the existence of knots in the star
formation region on a scale of 2-3 HST pixels corresponding to a spatial scale
of $1-1.5 \> \rm kpc$.

\subsection{The disruptive effect of the jet-cloud interaction}

It is well known (e.g. \citeN{klein94} and references therein) that shocks can disrupt clouds
on timescales of the order of a shock crossing timescale, $t_{\rm sh} \sim 9.7 \times
10^5 \, (L/{\rm kpc})(v_{\rm sh}/10^3 \> \rm km  \> s^{-1})^{-1} \> \rm yrs$ where $L$ is the
relevant scale-size. For the transverse size of $3.4 \> \rm kpc$ ($L = 1.7 \rm kpc$) this
would be of the order of $1.6 \times 10^6 \rm yrs$ if the transverse velocity is as high as
$1000 \> \rm km \> s^{-1}$. However, the transverse component of the bow-shock velocity is
less than the velocity of advance of the bow shock and the shock-shredding timescale is
likely to be at least a factor of two higher than this estimate. In this case the shock
shredding timescale would be comparable to or higher than the star formation timescale,
especially for a cloud density $\sim 10 \> cm^{-3}$. Another way of looking at this is that
on the radio source timescale $\sim 10^7 \> \rm yrs$, the bow shock should propagate to the
edge of the cloud. When that happens one does not expect much of that region of the cloud to
survive. From the HST images in Figure~1, this appears to be the case. The
emission-line and enhanced star-formation activity is confined to the arc-shaped region near
B2. 

\subsection{Relation to the dynamics of the radio source}

An important consistency check on any model for a radio source relates to the radio
luminosity and jet energy flux. In principle, one can use
estimates of the jet energy flux obtained from the monochromatic power,
$P_\nu$ of a lobe and an estimate of the ratio $\kappa_\nu$ of monochromatic
power to jet energy flux \cite{bicknell96c,bicknell98a}. These estimates
depend upon the age of the source and the estimate of the magnetic field in
the lobe. The application of this method to high redshift radio galaxies meets
with some complications resulting from the fact that the theory applies to the
``low frequency'' region of the spectrum defined to be that region for which
the frequency is less than the radiative break frequency. Moreover the
estimation of the minimum energy magnetic field also depends upon measurements from the low
frequency region which, generally for high redshift radio galaxies, is
inaccessible. We therefore adopt the following approach: The estimate of
$\kappa_\nu$ and the estimate of the minimum energy magnetic field are both
equally valid if applied to the {\em extrapolated} low frequency spectrum. 
At frequencies greater than the break frequency, both the measured
flux density and power of a particular component at frequencies $\sim 1
\> \rm GHz$ are underestimates of the extrapolated quantities. If a spectrum breaks at a
frequency $\nu_b$ with a change in spectral index of $\Delta \alpha$, then the ratio of
extrapolated to measured flux densities is
$(\nu/\nu_b)^{\Delta \alpha}$. If, as in the standard injection plus cooling
model, $\Delta
\alpha =0.5$, then a ratio of extrapolated to measure flux densities $\sim 10$
implies a  break frequency in the observer's frame $\sim 10 \> \rm MHz$ and a
corresponding break frequency in the rest frame $\sim 50 \> MHz$ . It is therefore unlikely
that the ratio of extrapolated to measured flux densities is greater then 10.

What do we take to be the ``lobe'' in 4C41.17? \shortciteN{chambers90a} have
argued that this should be component B3 rather than component C, which has a
very steep spectral index and does not appear to be connected to the inner
components. Component C is possibly a relic of earlier activity in
this galaxy and we adopt the \shortciteN{chambers90a} interpretation of
component B3 as the lobe. The $1.5 - 4.7 \> \rm GHz$ spectral index of B3 is 1.2
\cite{carilli94a}, consistent with a typical low frequency spectral index of
0.7 and a cooling induced break of $\Delta \alpha \approx 0.5$. Therefore the
above remarks on the extrapolated flux density are pertinent.

The monochromatic luminosity of B3 at rest frequency 
$\nu_{\rm rest} = (1+z) \nu_{\rm obs}$ is given by
\begin{equation} P_{\nu_{\rm rest}} = \frac {4 \pi D_{\rm L}^2}{1+z} \>
F_{\nu_{\rm obs}} 
\end{equation} where $D_{\rm L} = 5.16 \times 10^4 \> \rm Mpc$ is the
luminosity distance and $F_{1.465 \, \rm GHz} = 85 \> \rm mJy$
\cite{carilli94a}. Thus, $P_{7.0} \approx 5.7 \times 10^{34} \> \rm ergs \>
s^{-1}
\> \rm Hz^{-1}$. 

We estimate the ratio of monochromatic luminosity to jet power using
\begin{equation}
\kappa_{\nu} \approx (a-2) \, C_{\rm syn}(a) \: (\gamma_0 m_{\rm e}
c^2)^{(a-2)} \:
\left[ 1-(\gamma_1/\gamma_0)^{-(a-2)} \right]^{-1} \:   \,
 \: B^{(a+1)/2} \nu^{-(a-1)/2} \: \tau
\end{equation}  In this equation $a$ is the electron spectral index ($N(E)
\propto E^{-a}$), $\gamma_0$ and
$\gamma_1$ are the upper and lower cutoffs in the Lorentz factor of the
electron distribution,
$B$ is the magnetic field, 
$C_{\rm syn}(a) = 4 \pi c_5(a) c_9(a) (2c_1)^{(a-1)/2}$ incorporates a number of
\citeN{pacholczyk70} synchrotron parameters and $\tau = f_{\rm e} f_{\rm ad} t$ is an
evolutionary parameter which depends on $f_{\rm e}\approx 1$ the electron/positron fraction
of the internal energy, an adiabatic factor
$f_{\rm ad} \sim 0.5$ and the age of the lobe. In order to obtain theoretical
estimates of
$\kappa_{7.0}$ we estimate the minimum energy magnetic field from the peak
surface brightness of knot B3 in the $1.465 \> \rm GHz$ image
\cite{carilli94a} and a FWHM of
$0.11\arcsec$ \shortcite{chambers90a}. Since there are a number of uncertain
parameters, we bracket the estimates by values of extrapolated flux densities
and powers between 1 and 10 times the measured values and values of $\tau = 10 \rm Myr$.
(\shortciteN{chambers90a} estimated the dynamical
lifetime of B3 to be approximately $3 \times 10^7 \> \rm yr$ corresponding
to $\tau
\approx 15 \> \rm Myr$.) The other important parameter in these calculations is
the ratio of the actual value of the magnetic field to the minimum energy value. This is
probably the most important parameter for reconciliation of the radio power and the jet energy
flux, since $\kappa_\nu \propto B^{1+\alpha}$. For Cygnus~A,
\citeN{carilli91b} estimated that the
magnetic field strength is about 0.3 times the minimum energy value in order to reconcile
the lobe advance speed estimated from spectral aging with that estimated from ram pressure
balance. Likewise, \citeN{wellman97a} and \citeN{wellman97b}, using the same argument,
estimated that $ B \approx 0.25 B_{\rm min}$ from a sample of powerful radio galaxies.
Therefore, in table~\ref{t:jet-power}, the results of the calculations for
$\kappa_{7.0}$ and the resultant estimates of the energy flux,
$F_{\rm E} =
\kappa_{7.0}^{-1} \, P_{\rm 7.0}$ are given for different extrapolated flux densities,
a value of $\tau = 10 \rm Myr$ and for $B=B_{\rm in}$ and $B=0.25 B_{\rm min}$. As one can
see from the table, the radio power is consistent with a $\hbox{jet energy flux} \sim
10^{46} \> \rm ergs \> s^{-1}$ for $F_{\nu,\rm extrap} / F_{\nu} \ga 10^{0.5}$ and
$B/B_{\rm min} = 0.25$ and this is consistent with the upper limits on the jet energy flux
given by equations~\ref{e:fe}.

\subsection{Cloud mass and gravitational instability of clouds in the halo of
4C41.17}

In order to estimate the cloud mass involved in this interaction, we assume
that the cloud is as deep as it is wide (3.4~kpc) and we assume an area equal
to the extent of the entire \lya-bright region adjacent to B2 ($\approx 65 \>
\rm kpc^2$). This yields a mass $\approx 8 \times 10^{10} f({\rm \civ})
\msolar$.  The cloud mass inferred for the cloud interacting with the jet is
well in excess of the Jeans mass. Presumably this cloud is not exceptional as
far as clouds in the halo of 4C41.17 are concerned. Indeed, the existence of
other regions forming stars, albeit at a lower rate (see table~{t:sfr} and
\citeN{vanbreugel98a}), indicates that star formation is also occurring within
this galaxy as a result of other more standard processes ensuing from
gravitational collapse. The freefall time for a cloud of density
$\rho$ is $t_{\rm ff} \approx 2100 \> \rho^{-1/2} \approx 1.4 \times 10^7 \>
\left( n_{\rm H} /10 \>
\rm cm^{-3} \right)^{-1/2} \> \rm yr$ -- comparable to the dynamical timescale of
the radio source itself.

\subsection{Other regions of Ly$\alpha$ flux}

For a {\em uniform} density of $n_{\rm H} = 1.0 \rm cm^{-3}$, the extent of
the ionized precursor zone for a
$700-900\>\rm km\> s^{-1}$ shock  at LMC metallicity ranges from $4.2 - 9.4 \>
\rm kpc$  (Sutherland, Allen \& Kewley 1999). The extent of the precursor
emission is inversely  proportional to density. Hence, the 
\lya\ emitting region, approximately 4~kpc in extent to the east of B3 could
be shock precursor emission from a region of density $\sim 1.1-2.4 \>
\rm cm^{-3}$. Similarly the other \lya\ emitting region, $\sim 10 \> \rm kpc$
in extent on the western side of the galaxy and beginning at another radio
knot could be precursor emission associated with the expansion of that radio
component if the density is approximately $0.4-0.9 \> \rm cm^{-3}$. Thus, both of
these regions could be the result of jet interactions in slightly more tenuous regions.

\section {Conclusions}

We have shown that the recent observational evidence for jet-induced star
formation in 4C41.17 can be understood through a model in which clouds with a
density ($n_{\rm H} \sim 1-10 
\> \rm cm^{-3}$) in the galaxy are compressed by the bow shock resulting from
interaction with the jet.  Using the \civ\ emission to estimate the density
leads to a consistent scenario for shock induced star formation. The
gravitational timescales estimated from the density are well within the
dynamical timescale of the radio source and suggest that, compared to the
radio source timescale, star formation should occur almost instantaneously.
Some of the important features of the observations which lead to a reasonably
self-consistent model include the large velocities with respect to systemic
and large velocity dispersions of the \lya\ emitting region near B2. These
imply relatively large bow-shock velocity $\sim 1000 \> \rm km \> s^{-1}$ and
an upper limit on the jet energy flux $\sim 8 \times 10^{46} \rm ergs \>
s^{-1}$. These estimates are consistent with the radio source energy budget if the size of
the momentum deposition region is smaller than the upper limit $\sim 0.11 \arcsec$ and/or if
the magnetic field is approximately 0.25 times the equipartition value. Our analysis
therefore supports the general picture of jet-induced star formation in the papers cited in
the introduction.

In view of the wealth of data on 4C41.17 and the physics revealed by the
combined HST and VLA observations, it is clear that jet-induced star formation
can indeed be a significant process in many very high-redshift radio galaxies.
Theoretically this was anticipated 10 years ago, on the basis of the dense,
cloudy media expected in forming galaxies and the presence of extremely
luminous (10 - 100 times the luminosity of Cygnus A) radio sources embedded in
them.  Observationally, at present, 4C41.17 is  unique in that direct evidence
for jet-induced star formation exists. However, the radio/UV alignments seen
in many $z > 3$ radio galaxies, together with their blue colors and the very
similar rest-frame optical and radio source sizes
\cite{vanbreugel98a}strongly suggest that jet-induced star bursts may occur in
most high-redshift radio galaxies.

\acknowledgements
 
WvB appreciates the support of the Australian National University Astrophysical Theory
Centre, the Anglo-Australian Observatory, and the
Australian National Telescope Facility during his sabbatical leave at these
institutions from January to April, 1997.  He thanks his Australian colleagues
for their warm hospitality and invigorating discussions.  The work by WvB at
IGPP/LLNL is performed under the auspices of the US Department of Energy under
contract W--7405--ENG--48. AD acknowledges the support of NASA~HF-01089.01-97A
and partial support from a postdoctoral Research Fellowship at NOAO, operated
by AURA, Inc. under cooperative agreement with the NSF. We are grateful to the referee of the
original manuscript for detailed and helpful comments which have contributed significantly to
an improvement of the science presented.

\newpage


\newpage

\begin{center}
\large{\bf Tables}
\end{center}

\vskip 2.5 cm

\begin{table}[h]
\begin{center}
\begin{tabular}[]{c c c}
\hline \hline Component & Diameter & SFR \\
          &  (kpc)   & ($\msolar \> \rm y^{-1}$) \\
\hline NW        & 11       &   60 \\ NE        & 11       &   200 \\
NEE       & 11       &   30  \\ S         & 22       &  110  \\
\hline
\end{tabular}
\end{center}
\caption{Star formation rates in the different UV components, estimated by Van
Breugel et~al. (1998).}
\label{t:sfr}
\end{table}

\vskip 55 pt

\begin{table}[h]
\begin{center}
\begin{tabular}[]{c c c c c c }
\hline \hline
$F_\nu^{\rm extrap}/F_\nu^{\rm obs}$ & $P_{7.0 \rm GHz}$ & $B_{\rm min}$ &
$B/B_{\rm min}$ & $\kappa_{7.0}$ &
$F_{\rm E}$  \\
 & $\rm ergs \> s^{-1} \> Hz^{-1}$ & (Gauss) &  & $\rm (Hz^{-1})$ &
$\rm (ergs \> s^{-1})$  \\
\hline 1 & $5.7 \times 10^{34}$ & $3 \times 10^{-4}$ & $1$ & $1 \times
10^{-10}$ & $6 \times 10^{44}$ \\
  &                      &                    & $0.25$ & $9.5 \times 10^{-12}$ &
$6 \times 10^{45}$ \\
\hline
$10^{1/2}$& $1.8 \times 10^{35}$ & $3.5 \times 10^{-4}$ & $1$ & $2 \times
10^{-10}$ & $1 \times 10^{45}$ \\
          &                      &                      & $0.25$ & $2 \times
10^{-11}$ & $1 \times 10^{46}$ \\
\hline 10      & $5.7 \times 10^{35}$ & $5 \times 10^{-4}$& $1$ & $3 \times
10^{-10}$ & $2 \times 10^{45}$\\
   &                     &                    & $0.25$ & $3 \times 10^{-11}$ &
$2 \times 10^{46}$ \\
\hline
\end{tabular}
\end{center}
\caption{Estimates of the jet energy flux from the radio power of component
B3.}
\label{t:jet-power}
\end{table}

\newpage

\begin{center} {\Large \bf Figure Captions}
\end{center}

\begin{itemize}

\item[\bf Figure 1] Montage of three HST images taken through
the F702W, F569W and \lya\ filters. The X-band radio images of
\citeN{carilli94a} are superimposed in the form of contours. 

\item[\bf Figure 2] The F702W HST image smoothed to $0.2
\arcsec$ resolution. This enhances the star-forming region to the South.

\item[\bf Figure 3] Radiative shock models from \cite{sutherland99a} for 
a range of metallicities.  The shaded bar indicates the range of $\alpha(\civ)$
for shock velocities over $600\>\rm km \> s^{-1}$  The curves are labeled with the
metallicities of each series.  The straight line labeled DS95 is a least squares
fit to the results of the $150-500\>\rm km \> s^{-1}$  grid of solar metallicity
models from \citeN{dopita96a} and \citeN{dopita96b}.  The earlier grid does not
extrapolate well to the higher velocity range of $500-900\>\rm km \> s^{-1}$, where
the \civ\ emission from the precursor rapidly rises producing the 
higher than expected $\alpha(\civ)$ values.  Lower metallicity models with SMC and
LMC abundances are very efficient \civ\ producers due to hotter precursors in those models.

\item[\bf Figure 4:] The suggested morphology of a jet-cloud
interaction as it relates to 4C41.17. The deflection of the jet at the cloud
is mediated by a shock which is responsible for the knot of radio emission.
The high pressure drives a radiative shock into the cloud and this initiates
star formation. Some of the narrow emission lines ($\sigma \approx 220-270 \>
\rm km\> s^{-1}$) characteristically associated with star formation originate
from this region. Other narrow emission lines, in particular \civll, originate
from the shock-excited line emission from behind the radiative cloud shock.
The low Mach number flow behind the jet shock causes dense filaments of gas to
be drawn out of the cloud via the Kelvin-Helmholtz instability. This is the
origin of the broader ($\sigma \approx 470-600 \> \rm km \> s^{-1}$) emission
lines. 

\item[\bf Figure 5:] Illustration of a radiative shock and the
associated post-shock radiatively cooled layer wherein star formation is envisaged to
occur. As the shock progresses through the pre-shock gas (density $\rho$) the column
density of the cool layer accumulates rendering it more and more gravitationally unstable. 

\end{itemize}

\end{document}